\documentclass[aps,pre,reprint,groupedaddress]{revtex4-1}
\usepackage{amssymb}
\usepackage{amsmath}
\usepackage{color}
\usepackage{graphicx}
\usepackage{epsfig}
\usepackage{epstopdf}
\usepackage{dsfont}

\newcommand{\grl}{    {Geophys. Res. Lett.}}
\newcommand{\jgr}{    {J. Geophys. Res.}}
\newcommand{\ssr}{    {Space Sci. Rev.}}
\newcommand{\planss}{    {Plan. Sp. Sci.}}

\newcommand{\solphys}{ {Solar Physics}}

\newcommand{\apjl}{    {Astrophys. J. Lett.}}

\newcommand{\Ham}{\mathcal{H}}

\begin{document}


\title{Regimes of charged particle dynamics in current sheets: the machine learning approach}



\author{Lukin A.S.$^{1,2}$}
\email[]{as.lukin.phys@gmail.com}

\author{Artemyev A.V.$^{3,1}$}

\author{Vainchtein D.L. $^{4,1}$}

\author{Petrukovich A.A.$^{1}$}


\affiliation{
$^1$ Space Research Institute RAS, Moscow, Russia;
$^2$ Faculty of Physics, National Research University Higher School of
Economics, Moscow, Russia;
$^3$ Institute of Geophysics and Planetary Physics, University of California, Los Angeles, CA, USA;
$^4$ Nyheim Plasma Institute, Drexel University, Camden, NJ, USA;}


\date{\today}

\begin{abstract}
Current sheets are spatially localized almost-1D structures with intense plasma currents. They play a key role in storing the magnetic field energy and they separate different plasma populations in planetary magnetospheres, the solar wind, and the solar corona. Current sheets are primary regions for the magnetic field line reconnection responsible for plasma heating and charged particle acceleration. One of the most interesting and widely observed type of 1D current sheets is the rotational discontinuity, that can be force-free or include plasma compression. Theoretical models of such 1D current sheets are based on the assumption of adiabatic motion of ions, i.e. ion adiabatic invariants are conserved. We focus on three current sheet configurations, widely observed in the Earth magnetopause and magnetotail and in the near-Earth solar wind. Magnetic field in such current sheets is supported by currents carried by transient ions, which exist only when there is a sufficient number of invariants. In this paper, we apply a novel machine learning approach, AI Poincar\'{e}, to determine parametrical domains where adiabatic invariants are conserved. For all three current sheet configurations, these domains are quite narrow and do not cover the entire parametrical range of observed current sheets. We discuss possible interpretation of obtained results indicating that 1D current sheets are dynamical rather than static plasma equilibria.
\end{abstract}


\maketitle
\section{Introduction}

Current sheets are a common feature of space plasma systems. They are observed in planetary magnetospheres \cite{Jackman14}, the solar wind \cite{Tsurutani&Ho99,Gosling12}, and the solar corona \cite{bookParker94,bookBirn&Priest07}. Current sheets are characterized by spatially localized plane (surface) current density supporting strong magnetic field gradients. Depending on the value of plasma $\beta$ (ratio of thermal plasma pressure and magnetic field pressure), current sheets may contain mostly diamagnetic cross-field currents (large-$\beta$ systems, like planetary magnetotails \cite{Petrukovich15:ssr,Artemyev14:pss,Halekas06:currentsheets,DiBraccio15, Dubinin12,Poh17}) or mostly field-aligned currents (low-$\beta$ systems, like the solar wind \cite{Vasquez07,Greco09,Greco16,Vasko21:apjl}). In some space plasma systems, depending on external conditions, both large-$\beta$ and small-$\beta$ regimes may occur. The current sheet with cross-field currents are compressional plasma slabs, whereas current sheets with purely field-aligned currents are force-free compressionless slabs. An example is the planetary magnetopause, a boundary between the solar wind flow and planetary magnetosphere, where various $\beta$ regimes are observed \cite{deKeyser05,Panov08,Haaland19,Haaland20,Lukin20}. Current sheets play an important role in the magnetic field energy dissipation leading to plasma heating and charged particle acceleration \cite{Paschmann13,Sitnov19,Servidio15}. Magnetic field-line reconnection  is a key process for such dissipation   \cite{bookPriestForbes00:reconnection,Birn&Hesse09,Servidio11} and releases of the stored magnetic field energy. The onset of magnetic reconnection and details of particle acceleration strongly depend on magnetic field in stationary current sheets \cite{Sitnov21:grl,Runov21:jastp,Phan18:nature,Lu20:natcom}. Such configurations are solutions of a stationary Vlasov-Maxwell system describing a self-consistent magnetic field and charged particle distributions \cite{bookVedenyapin11}. Stationary solutions are based on particle distribution functions written in terms of particle integrals of motion \cite[e.g.,][]{Artemyev&Zelenyi13, Sitnov&Merkin16}, (see also reviews of current sheet models in \cite{Priest85,bookParker94,Roth96,Sitnov19}). In 1D stationary systems, where energy and two canonical momenta are conserved, there are sufficient number of such integrals (see examples of 1D current sheet models for solar wind and magnetopause in  \cite{deKeyser96,deKeyser13,Panov11:magnetopause,Harrison09:prl,Allanson17:grl}). In 2D systems (where the energy and one canonical momentum are conserved), stationary current sheets without field-aligned currents can be constructed (see examples of 2D current sheet models for planetary magnetotails in \cite{Birn04,YL05,Catapano15,Sitnov&Merkin16}). However, such 1D and 2D current sheets describe tangential discontinuities, the most unstable \cite{Loureiro07,Bhattacharjee09,Pucci&Velli14,Pucci18} and less observed \cite[see discussion in][]{Neugebauer06,Artemyev19:jgr:solarwind} type of current sheets. The most interesting in context of the solar wind \cite{Medvedev97:pop,Vasquez&Hollweg99,Servidio15} and planetary magnetospheres \cite{Sitnov06,Zelenyi11PPR,Artemyev11:pop} are rotational discontinuities with or without field-aligned currents. They  do not possess a sufficient number of {\it exact} integrals of motion to construct a self-consistent spatially localized plasma equilibrium. Thus, models of such current sheets often include additional {\it adiabatic} invariants, approximately conserved for specific magnetic fields \cite[see discussion in][]{Sonnerup71,Whipple84,Whipple86,BZ89,Zelenyi13:UFN}. Therefore, to determine the number of adiabatic invariants is quite important for current sheet model construction.

The presence of adiabatic invariants is a consequence of a quasi-periodical motion of charged particles \cite{bookLL:mech}. For strong magnetic fields, such motion is naturally decomposed into three periodical motions with corresponding adiabatic invariants: magnetic moment for gyrorotation, the second adiabatic invariant for bounce oscillations in magnetic field traps, and the third adiabatic invariant for azimuthal drift (in time-dependent systems with closed drift paths) \cite[e.g.,][]{bookLyons&Williams,bookSchulz&anzerotti74,Ukhorskiy&Sitnov13:ssr}. Situation is more complicated in current sheets where the separation of different time-scales is not so pronounced \cite[see][]{Vainchtein05,Zelenyi13:UFN,Malara21}. For many concrete configurations it is not known if there are adiabatic invariants. Recently a new approach for the invariant search based on machine learning was proposed \cite{Liu&Tegmark21:ML}. In this study we apply this approach to three current sheets typical for the Earth’s magnetopause/magnetotail and the solar wind to check number of invariants for different parameters. 

The structure of the paper is as follows. First, in Sect.~\ref{sec:dynamical_systems}, we discuss several relevant aspects of the dynamical systems theory, in particular several time-scales that describe systems and their simulations. In Sect.~\ref{sec:approach}, we describe the adaptation of a machine learning approach from \cite{Liu&Tegmark21:ML} to charged particle dynamics in current sheets. Then we describe results of application of this approach for three current sheets in Sect.~\ref{sec:examples}. And finally we discuss obtained results and formulate main conclusions of this study in Sect.~\ref{sec:discussion}.


















\section{Dynamical systems}\label{sec:dynamical_systems}
We consider Hamiltonian functions $H(\bold{X},\bold{\lambda})$, where vector $\bold{X}=[\bold{x},\bold{p}]$ consists of particle's coordinates $\bold{x}$ and momenta $\bold{p}$, and $\bold{\lambda}$ is the vector of system parameters, assumed to be constant. Since $H(\bold{X},\bold{\lambda})$ doesn't depend on time explicitly, the total energy $H$ is an integral of motion. Our goal is to determine whether there are additional integrals of motion. 

In our work we are searching for both exact and adiabatic invariants. Adiabatic invariants are features of multi-scale quasi-periodic systems. They are the quantities approximately conserved in the process of motion. The accuracy of their conservation (or, equivalently, the rate of destruction) depends on the ratio of the timescales of quasi-periodic motions constituting a multi-scale system. The rate of destruction defines the time over which an adiabatic invariant can be assumed to be (approximately) conserved, $T_{AI}$. This time scale should be compared with two other time scales: a residence time $T_R$ of a given particle (phase point) in the system, and the lifetime $T_{syst}$ of the system itself. In well-defined multi-scale quasi-periodic systems it is assumed that all these time scales are much longer than the longest of the (quasi)-periods. In the presence of adiabatic invariants, the effective number of conserving quantities detected in any numerical simulations depends on timescale of simulation, $T_{Sim}$. In our work we are interested only in total number of integrals conserving on time scales $T_{Sim} \sim T_{Syst}$, and we make no difference between the exact and adiabatic ones. However, AI Poincar\'{e} \cite{Liu&Tegmark21:ML}, the method described below, can also be used to determine time-scales of approximate invariants conservation.

For a dynamical system with a state vector $\bold{x} \in \mathds{R}^N$, the phase space of $\bold{X}$ is $D=2N$-dimensional. Each conservation law (corresponding to an invariant of motion) reduces the number of independent variables. If there are $K$ independent integrals of motion, then all particle’s trajectories lie on manifolds $\mathcal{M}$ with a dimension $M=D-K$. Thus, the problem of determining the number of conservation laws (integrals of motion) can be reduced to the problem of estimating of dimensionality of the manifold $\mathcal{M}$.

\begin{figure}
\centering
\includegraphics[width=0.22\textwidth]{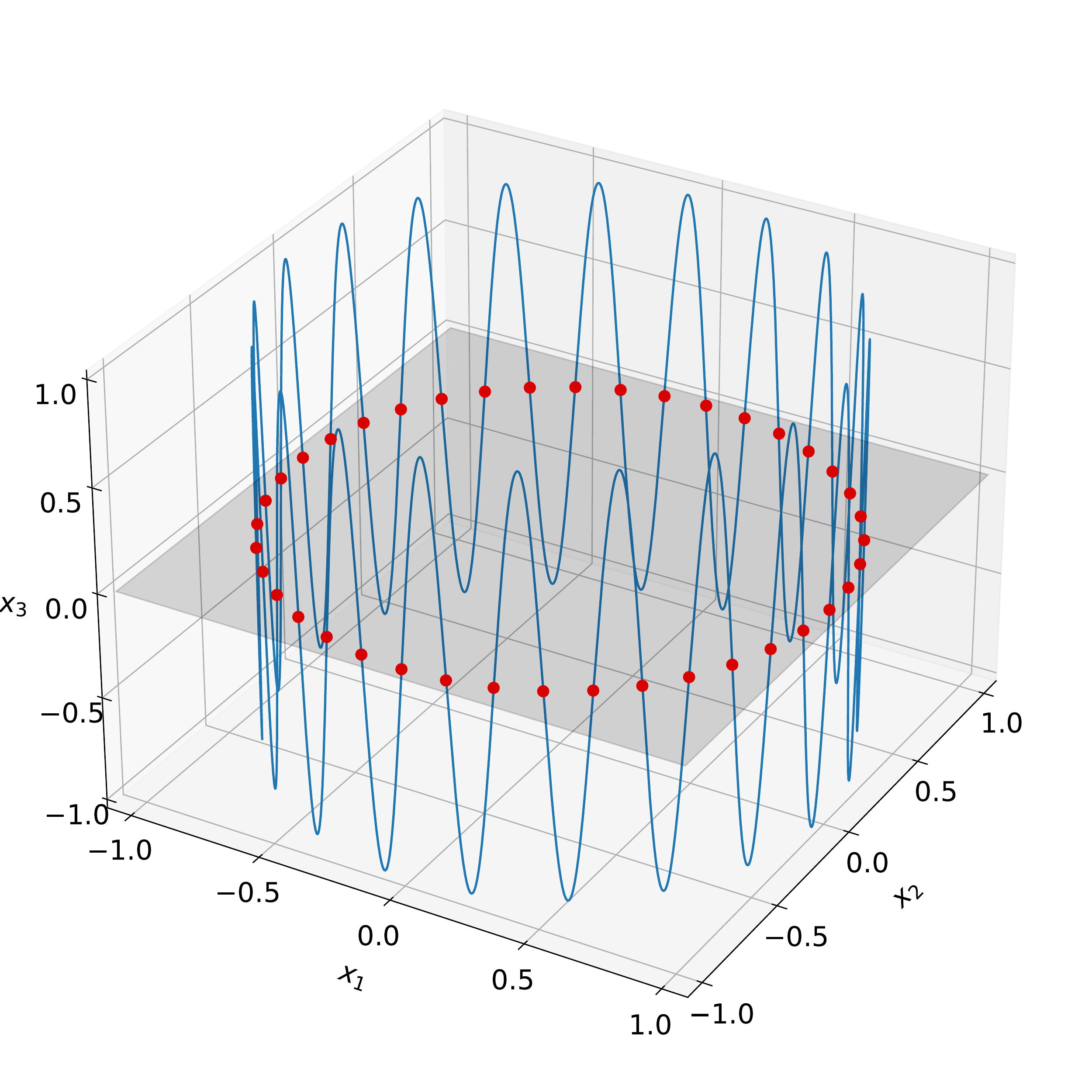}
\caption{
\label{poinkare_map} Poincar\'{e} map for an artificial system. The motion are periodic in all three dimensions. Particle trajectory in a 3D space (blue), Poincar\'{e} surface of section, $x_3=0$ plane, (shaded gray), and Poincar\'{e} cross section (points of intersection of the trajectory with section plane) (red dots).}
\end{figure}

Sometimes it is more convenient to study not the particle trajectory itself, but its Poincar\'{e} map (see a schematics in Fig.~\ref{poinkare_map}). The dimensionality of the Poincar\'{e} map is $D=2N-1$. Trajectories of all systems in this study quasi-periodically cross a certain  plane $x_j=const$. If a trajectory is on an embedded manifold $\mathcal{M}$, the dimension of its Poincar\'{e} map is $M=2N-1-K$. From the dimensionality of the Poincar\'{e} maps we can estimate the number of conserving quantities $K$.


\section{The machine learning approach}\label{sec:approach}

Several machine learning approaches have been proposed for searching integrals of motion \cite{Liu&Tegmark21:ML,Liu&Tegmark22:Poinkare2,Mototake21:IntegralsNN,Wetzel20:SNN}. They are mostly based on the assumption that if additional invariants exist, the particle trajectories reside on a lower-dimensional manifold embedded in a phase space \cite[e.g.,][]{Cayton05:Manifolds,VanDerMaaten08:t-SNE,Izenman12:Manifolds}. One of the proposed approaches is to estimate $M$ (and thereby $K$) directly from numerical trajectories by using a nonlinear generalization of the principal component analysis (PCA) \cite[see e.g.,][]{Jolliffe16:PCA}. 

\subsection{Principal component analysis (PCA)}

Classical PCA is a linear method widely used for dimensionality reduction of datasets. It is an affine transformation of a multidimensional data set with $D$ variables onto a new orthogonal coordinate system of principal components (PCs). The PCs are constructed consecutively, each new principal component $PC_{j+1}$ must be orthogonal to all previous $PC_{k}$, $k=1,…,j$, and must maximize the variance $var_{j+1}$ of the original data projected onto it. The PCA is characterized by the explained variance ratios (or explained ratios) of the PCs: $\sigma_j={var_j}/\left(\sum_{k=1}^{D} var_k\right)$. Each subsequent PC has a smaller value of $\sigma_j$ and if $\sigma_j=0$ for some $j$, it means that the data are constant along $PC_{j}$. i.e. we can exclude all PCs starting from the $j$-th without a loss of information. In realistic applications, there is a finite threshold $\sigma_{min}>0$ (although its definition is somewhat heuristic): if $\sigma_j \leq \sigma_{min}$ for some $j$, the real dimensionality of the dataset is $M=\min(j)-1$. In this case there is an $M$-dimensional manifold $\mathcal{M}$ ($M<D$) embedded into the $D$-dimensional dataset.

In most of applications, if an embedded manifold exists, it is not a hyperplane. A direct application of classical PCA on the whole dataset can lead to an incorrect estimate of $M$. For example, for the trajectory in Fig.~\ref{poinkare_map}, a direct use of PCA to the whole data gives almost equal values of explained ratios $\sigma_j\sim 0.33$, because all variables $x_1,x_2,x_3$ have the same variations. In this case an estimate of the manifold dimensionality is $M=3$. However, as the points lie on a smooth curve, the real value is $M=1$. 

To estimate the dimensionality of a non-planar $\mathcal{M}$, we start by obtaining a large dataset $\bold{X}$. In our work, $\bold{X}$ is a Poincar\'{e} map of numerically integrated particle trajectories (see Sect.~\ref{subsec:datasets}). Then we apply PCA locally, in a small vicinity of any particular point $\bold{X}^*$. For that we need sufficiently many  points near $\bold{X}^*$. To sufficiently populate a vicinity of $\bold{X}^*$, we need to integrate particle trajectories over long time interval(s) or to use proportionally many initial conditions, which may take a long time and CPU-intensive. Alternatively, we can use a neural network (NN) to create a ``cloud'' of data points near $\bold{X}^*$. After that, the local PCA is performed at some (randomly selected) points $\bold{X}_i$, $k=1,...,10$, on the dataset. After that, the average values of the explained variance ratios are computed: $\hat{\sigma}_j=(1/N_s) \sum_{i=1}^{N_{s}}\sigma_j^i$, where $\sigma_j^i$ are computed near $\bold{X}_i$. The value of $M$ is then defined as a number of explained ratios $\hat{\sigma}_j$ which are greater than a predefined threshold $\sigma_{min}$.

\subsection{AI Poincar\'{e}}

\begin{figure}
\centering
\includegraphics[width=0.45\textwidth]{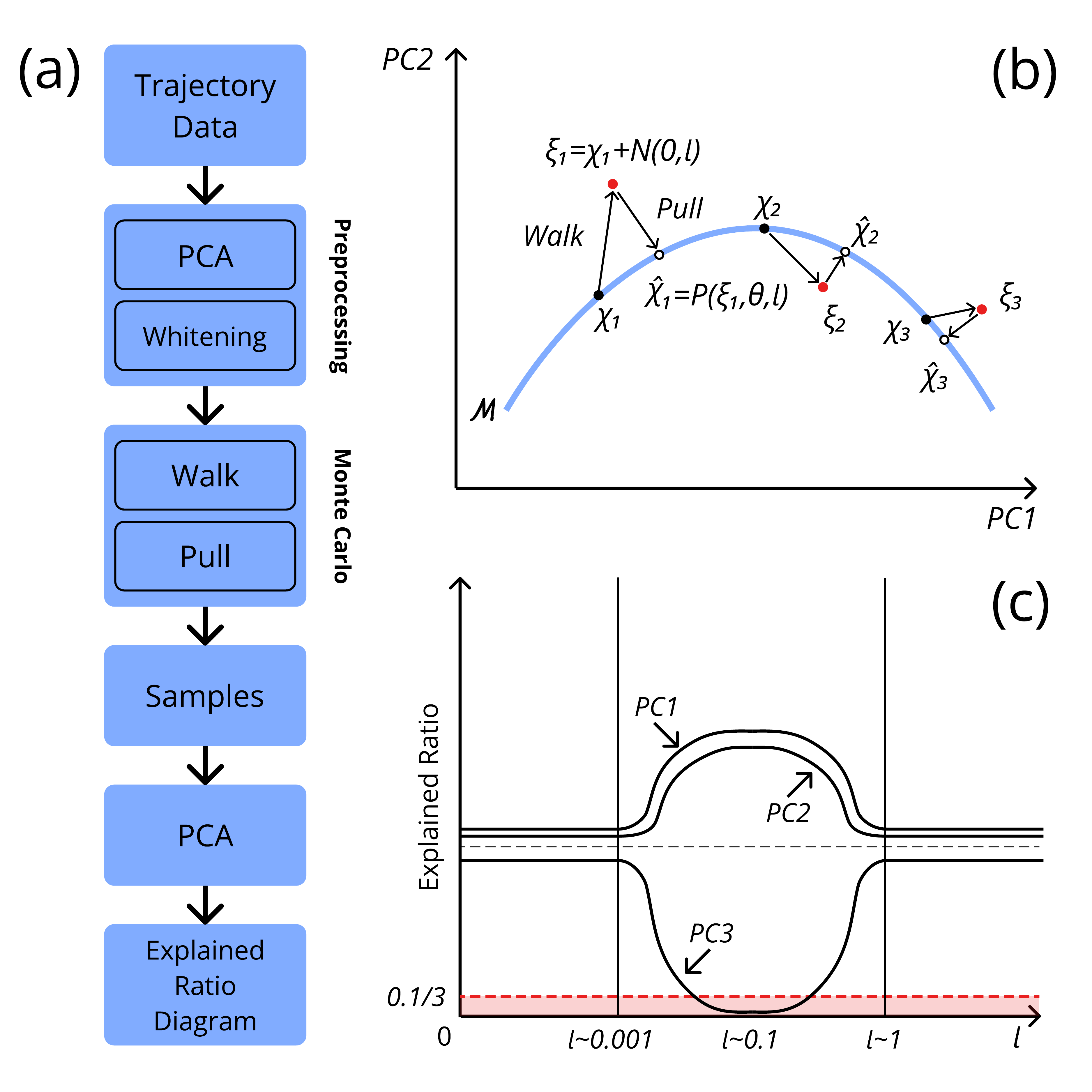}
\caption{
\label{scheme} A schematic diagram of AI Poincar\'{e} (the same as in Fig.1 in \cite{Liu&Tegmark21:ML}). (a) the method's pipeline. (b) schematic of the Monte Carlo sampling module. (c) an example of the explained ratio diagram with one identified conservation law.}
\end{figure}

AI Poincar\'{e} is a recently developed machine learning approach for determining of a total number of conservation laws \cite{Liu&Tegmark21:ML}. Main advantages of this approach are that it is fully automatic and requires only a dataset of numerically integrated trajectories. AI Poincar\'{e} can be used to estimate the dimensionality of any data subset, including the Poincar\'{e} maps. The pipeline of AI Poincar\'{e} consists of three modules (see Fig.~\ref{scheme}(a) or Fig.~1 in \cite{Liu&Tegmark21:ML}):
\begin{enumerate}
    \item Preprocessing/Whitening
    \item Monte Carlo sampling
    \item Linear dimensionality estimation
\end{enumerate}

\begin{figure}
\centering
\includegraphics[width=0.45\textwidth]{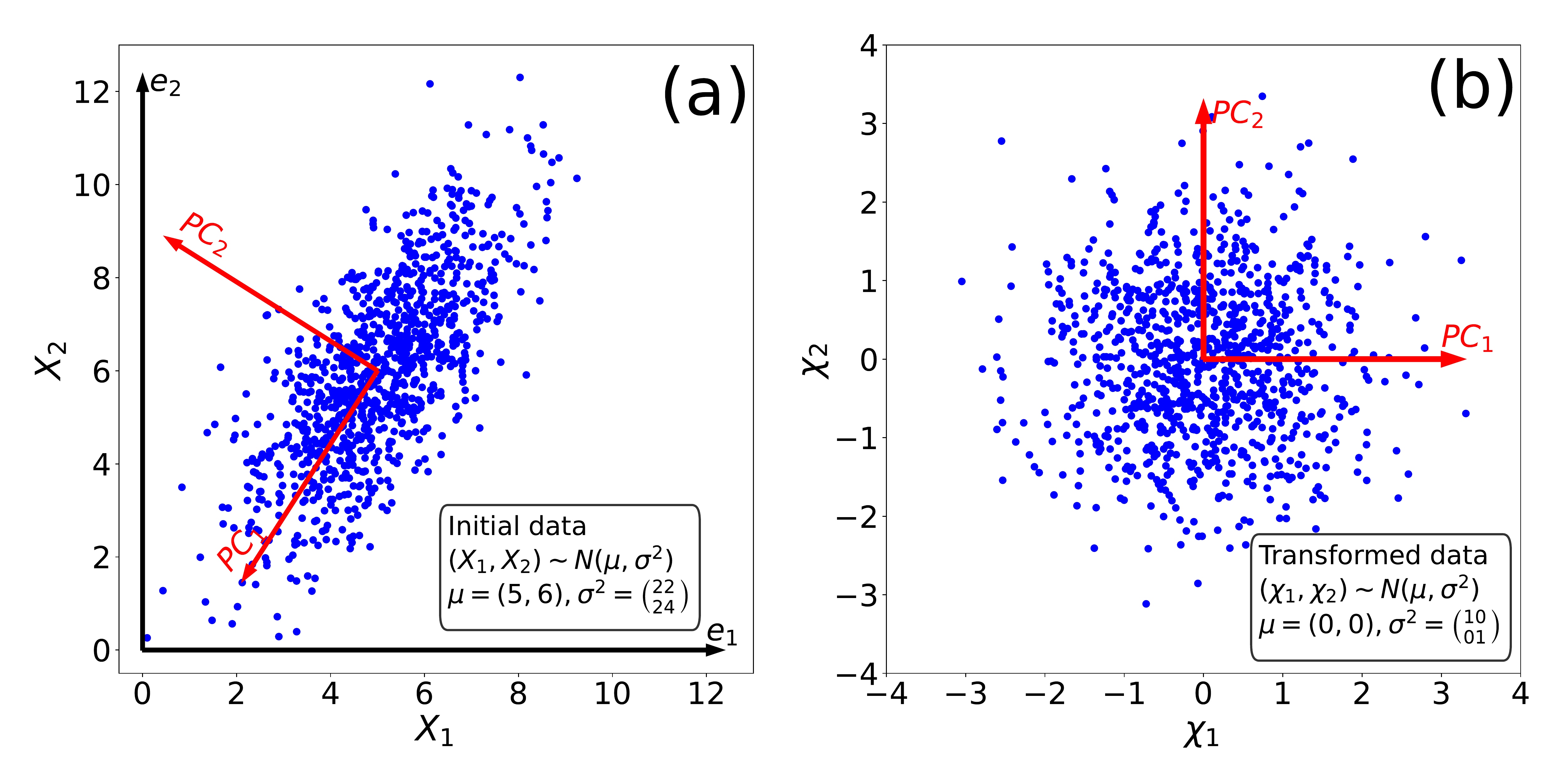}
\caption{
\label{PCA_example} An example of the PCA procedure with whitening for a two-dimensional dataset. (a): a scatter plot of the the original coordinates $(X_1,X_2)\sim N(\mu,\sigma^2)$ with $\mu=(5,6)$, $\sigma^2=\binom{2 2}{2 4}$ and the orientation of PCs. (b): transformed variables $(\chi_1, \chi_2) \sim N(\mu,\sigma^2)$ with $\mu=(0,0)$, $\sigma^2=\binom{1 0}{0 1}$.}
\end{figure}

AI Poincar\'{e} can optionally include the whitening of the variables \cite{Jolliffe16:PCA}. On the whitening step, we use a PCA to transform the coordinates and momenta into a new orthogonal basis $\bold{X}\rightarrow \boldsymbol{\chi}$ such that the new variables $\boldsymbol{\chi}$ have zero means and their covariance matrix is the identity matrix, i.e. the scales of all new variables are equal to one. Whitening is applied to the whole dataset $\bold{X}$. Figure \ref{PCA_example} shows an example of whitening for a two-dimensional dataset $(X_1,X_2)$. Panel (a) shows two PCs in the original coordinate system. Panel (b) shows a scatter plot of new variables $(\chi_1,\chi_2)$, transformed to the PCA coordinate system and whitened. This step is optional and the method works without it, but whitening simplifies the next module, when we train the neural network, because the scales of all the coordinates in $\boldsymbol{\chi}$ are equal.

The Monte Carlo sampling module is used to obtain new state vectors $\boldsymbol{\hat{\chi}_i}$ which lie close to the trajectory in the vicinity of a known vector $\boldsymbol{\chi_i}$ lying on the trajectory. This procedure is performed in two steps (see Fig.~\ref{scheme}(b)):
\begin{enumerate}
    \item Walk step: $\boldsymbol{\chi} \rightarrow \boldsymbol{\xi} = \boldsymbol{\chi} + \bold{n}$. We perturb a state vector by adding an isotropic Gaussian noise $\bold{n}$ with a zero mean and covariance matrix $l^2\bold{I}$, i.e. $\bold{n}\sim N(0,l^2\bold{I})$, where $l$ is the noise scale. 
    \item Pull step: $\boldsymbol{\xi}\rightarrow \boldsymbol{\hat{\chi}}$. We map the perturbed vector back towards the manifold.
\end{enumerate}

For the pull step we construct a mapping operator $P: P(\boldsymbol{\xi},\boldsymbol{\theta}, l)=\boldsymbol{\hat{\chi}}$, parameterized as a feed-forward neural network (NN) with $4$ hidden layers of $256$ neurons each with ReLU activation functions \cite{Glorot11:ReLu}. Parameters $\boldsymbol{\theta}$ are weights and biases of the neurons and are determined by optimizing the MSE Loss function using the Adam optimizer \cite{Kingma14:adam}:
\begin{equation}\label{MSELoss}
\begin{cases}
    \text{Loss}(\boldsymbol{\hat{\theta}},l)=\frac{1}{N_{train}}\sum_{i=1}^{N_{train}}|P(\boldsymbol{\xi_i},\boldsymbol{\hat{\theta}},l)-\boldsymbol{\chi_i}|^2_2 \\ \\
        \boldsymbol{\theta}=\text{argmin}_{\boldsymbol{\hat{\theta}}} \left( \text{Loss} \left(\boldsymbol{\hat{\theta}},l\right)\right)
\end{cases}
\end{equation}
where $N_{train}$ is the number of samples for training.

For training and validation, we start with \emph{Poincar\'{e} map dataset} consisting of $N_c=10,000$ state vectors $\boldsymbol{\chi}$, that were transformed to the PCA basis on the preprocessing step. We perturbed each sample vector $\boldsymbol{\chi_i}$ from the \emph{Poincar\'{e} map dataset} with $10$ different random vectors: $\boldsymbol{\xi}_i^k=\boldsymbol{\chi}_i + \bold{n}_i^k$, $k=1,\ldots, 10$ (Walk step). The NN takes the perturbed vectors $\boldsymbol{\xi}$ as an input while the unperturbed vectors $\boldsymbol{\chi}$ serve as a targets, i.e. NN tries to denoise the perturbed vectors (Pull step). The \emph{training/validation dataset} contains $N_{total}=100,000$ pairs of vectors $\boldsymbol{\xi}$. This dataset is randomly divided into training and validation subsets of $N_{train}=70,000$ and $N_{valid}=30,000$ vectors (this is a typical proportion most often used in machine learning) and used to train the mapping operator $P(\boldsymbol{\xi},\boldsymbol{\theta},l)$. With many training samples, NN learns the general shape of the trajectory (see schematics in Fig.~\ref{scheme}(b)). In the optimum scenario, the NN is trained to orthogonally project $\boldsymbol{\xi}$ back onto the trajectory. Validation subset is used to control the overfitting of the NN: we stop the training if the loss function on the validation subset doesn't change for $50$ epochs.

The main hyperparameter of this algorithm is the noise scale $l$. We need to train a separate NN for each value of $l$. When $l$ is large ($l \gtrsim 1$) or very small ($l \lesssim 10^{-3}$), we cannot correctly estimate $M$ \cite{Liu&Tegmark21:ML}. As the scales of $\bold{\chi}$ are of order one, so in the first case the noise has about the same scale as the variables and projected points are distributed all over the manifold. In the second case, the noise scale is too small and all the projected points lie too close. The optimal value of $l$ usually is around $l\sim 0.1$, so we trained mapping operators for $l \in [10^{-2}, 0.3]$. If $l$ is appropriately chosen and NN training is successful, the mapped points $\boldsymbol{\hat{\chi}}=P(\boldsymbol{\xi},\boldsymbol{\theta},l)$ are close to the manifold and describe its local tangent space. 

\subsection{PCA and estimating the dimensionality}

Once the mapping operator $P(\boldsymbol{\xi},\boldsymbol{\theta}, l)$ is obtained, we can perform the local PCA in different segments of the manifold. We randomly selected $N_s=2,000$ points $\boldsymbol{\chi}_i$ from the \emph{Poincar\'{e} map dataset}. This number is rather arbitrary and barely affects the speed of the algorithm; sufficiently large number of points smooths out potential errors in determining the local dimensionality of the manifold. At each point $\boldsymbol{\chi}_i$ we did $N_{wp}=10,000$ walk-pull steps $\boldsymbol{\chi}_i \rightarrow \boldsymbol{\xi}_i^k = \boldsymbol{\chi}_i+ \bold{n}_i^k \rightarrow \boldsymbol{\hat{\chi}}_i^k = P(\boldsymbol{\xi}_i^k,\boldsymbol{\theta}, l)$, $k=1...N_{wp}$. Then we performed PCA to the set of $\boldsymbol{\hat{\chi}}_i^k$, fixed $i$ and $k=1...N_{wp}$, to get the values of explained variance ratios $\sigma_j^i(l)$,  $j=1...D$. The bases of local PCA may be different at different $\boldsymbol{\chi}_i$. Each $\sigma_j^i(l)$ gives us a local estimate of the dimensionality in the vicinity of $\boldsymbol{\chi}_i$. Then we  averaged the explained ratios to get $\hat{\sigma}_j(l)=(1/N_s)\sum_{i=1}^{N_s} \sigma_j^i(l)$. Note, that $\hat{\sigma}_j$ can differ for different values of the noise scale $l$.


The output of the algorithm, the explained ratio diagram, allows us to estimate the dimensionality of the manifold and, thus, the number of conserving quantities. Figure~\ref{scheme}(c) schematically shows the averaged values of explained ratios $\hat{\sigma}_j(l)$ as a function of the noise scale $l$. Following \cite{Liu&Tegmark21:ML}, we set the threshold at $\sigma_{min}=0.1/D$, where $D=2N-1$ for the Poincar\'{e} map and $N$ is a dimension of state vector $\bold{x}$ (in our work $N=2$ and thus $\sigma_{min}=0.1/3$). If for some $l$, $\hat{\sigma}_j(l)<\sigma_{min}$, we assume that the data is constant along the $j$-th principal component. Thus the dimensionality of the manifold $\mathcal{M}$ can be estimated as
\begin{equation}
M=\text{min} (M_l), \quad M_l=\min{(j)}-1: \hat{\sigma}_j(l) \leq \sigma_{min}
\end{equation}
and thus $K=2N-1-M$.


\subsection{Datasets}\label{subsec:datasets}

In the present paper we consider three dynamical systems with Hamiltonians depending on two coordinates $(x_1,x_2)$ and two conjugate momenta $(p_1,p_2)$. Trajectories regularly cross the plane $x_2=0$. Therefore, instead of using trajectories themselves, we use Poincar\'{e} cross sections $\{x_1,p_1,p_2\}|_{x_2=0}$ to reduce the NN training time.

We integrated trajectories using the $4^{th}$ order Runge-Kutta scheme with non-dimensional $\Delta t = 5 \cdot 10^{-4}$. Each integration started from some initial point and we attempted to accumulate $N_c=10,000$ crossings of the $x_2=0$ plane. If/when a particle moved too far from the plane $x_2=0$ after however many crossings, we took a new initial point, satisfying all the requirements as the original initial point. and continued accumulating crossings until $N_c$ crossings are collected.  After the preprocessing step of AI Poincar\'{e} all the state vectors are transformed to the PCA basis $\{x_1,p_1,p_2\} \rightarrow \{\chi_1,\chi_2,\chi_3\}$ to get the \emph{Poincar\'{e} map dataset}.

$10,000$ crossings does not correspond to any concrete value of time. In different systems, and for different initial conditions and parameter values, $10,000$ crossings of the $x_2=0$ plane are accomplished in very different times. The results presented in the rest of the paper indicate the number of invariants over whatever time it takes to accumulate $10,000$ crossings. 

\subsection{Computational times}
A typical time required to accumulate 10,000 $z=0$ crossings is 5-10 minutes for one set of system parameters. A typical time of one NN training in our work is several minutes on AMD Ryzen 5900x processor and Nvidia RTX 3060 ti graphic card. The time needed to train the projector operator for one particular $l$ is about two minutes. The first operation can be efficiently parallelized and we used $24$ threads of AMD Ryzen 5900x processor. If we find the appropriate noise scale for a given Hamiltonian and certain set of parameters, for the nearby systems we can start training the projector operator for the same value of $l$. Most probably, that, or a close value of $l$ would be valid. For the figures in following sections we used $600$ different combinations of system parameters and the total time of computations for one system (particular view of Hamiltonian function) is about 24 hours. This time can be reduced by using more powerful CPU and graphic cards.

\section{Ion dynamics in current sheets} \label{sec:examples}
Current sheets observed in the planetary magnetotail, magnetopause, and solar wind are characterized by a reversal of the main magnetic field component, $B_l$, across a neutral plane, $r_n=0$. A standard approximation of this field is $B_l\approx B_0\cdot(r_n/L)$ where $B_0$ and $L$ are characteristic magnetic field magnitude and the current sheet thickness. The ratio $B_0/L$ is typically determined in spacecraft observations from the current density $j_m$ measurements: $B_0/L=4\pi\max j_m/c$ \cite{Artemyev16:jgr:pressure,Vasko15:jgr:cs}. The presence of a normal magnetic field component, $B_n \approx const$, makes magnetic field line curved and allows particles to cross the neutral plane. Separation of magnetic field magnitudes $B_0$ and $B_n$ defines what part of particle dynamics is fast and what part is slow. When $B_0\gg B_n$, which is typical for 1D current sheets \citep{bookSchindler06}, oscillations of $r_n$ are fast and of $r_l$ are slow. Thus there is an adiabatic invariant corresponding to the averaging over fast oscillations \citep{Schindler65,Sonnerup71,BZ86}. When $B_n \gg B_0$, the magnetic field reversal $B_l$ is embedded in a strong background $B_n$ field. In such systems the conserved adiabatic invariant is the magnetic moment \cite{bookSivukhin65,bookAlfven,bookNorthrop63}. These adiabatic invariants are widely used to construct current sheet models  \citep[e.g.,][]{Sitnov00,Zelenyi00,Sitnov06,Artemyev11:pop,Zelenyi11PPR}. They supplement the conservation of the total energy (in stationary current sheet models) and generalized momentum $p_{m}$ (in current sheet with $B_n\ne 0$ and 1D inhomogeneity coordinate, $r_n$). 

The number of invariants (conserving quantities) is the most important characteristic for construction of the kinetic plasma equilibria describing the current sheets. A velocity distribution function defined as a function of invariants is a solution of the stationary Vlasov equation. Thus the duration of conservation of adiabatic invariants. $T_{AI}$, determines the life-time $T_{syst, model}$ of the plasma equilibrium described by such model. A comparison of the predicted life-time $T_{syst, model}$ of the model with observational life-time $T_{syst}$ of current sheets can be used to verify the model applicability.

In the absence of adiabatic invariants (or when they are destroyed over the residence time of a particular particle in the current sheet, $T_{AI} < T_{R}$) the class of possible equilibrium current sheet models would be reduced to 2D solutions of Grad-Shafranov equation \citep[e.g.,][]{Kan73,YL05,Vasko13:pop}. 

\begin{figure*}[!]
\centering
\includegraphics[width=0.9\linewidth]{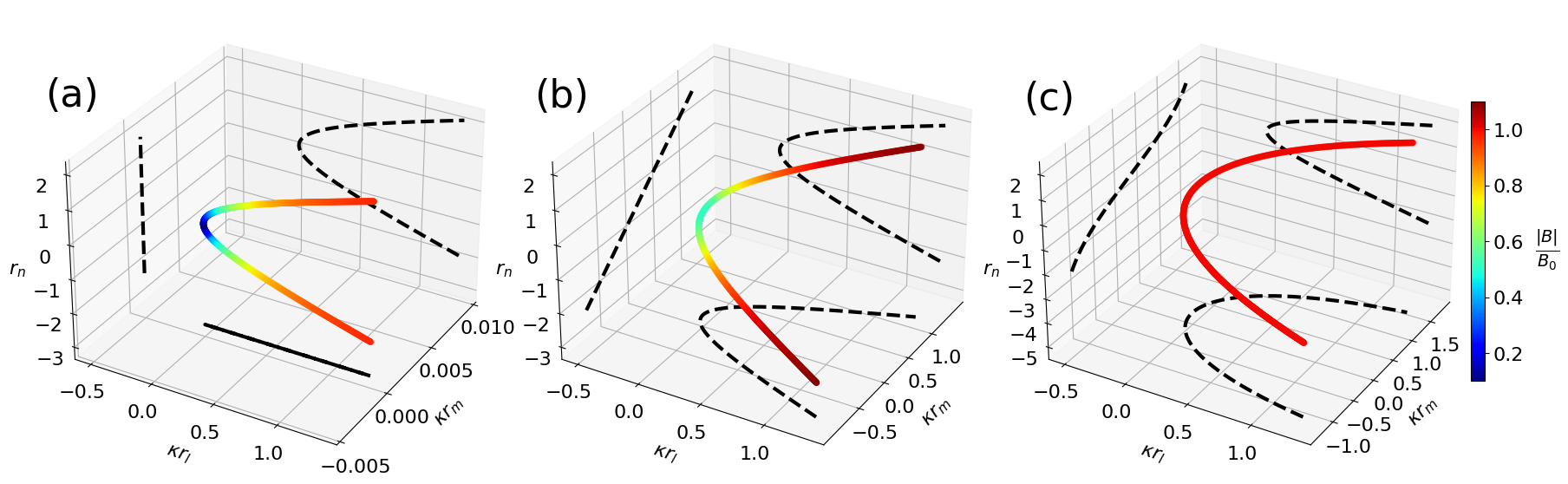}
\caption{
\label{fig1} Three current sheets: magnetic field-lines for current sheets with (a) $B_l(r_n)$, $B_n\ne 0$ and $B_m=0$, (b) $B_l(r_n)$, $B_n\ne 0$ and constant $B_m\ne 0$, (c) $B_l(r_n)$, $B_n\ne 0$ and $B_m(r_n)$. Color indicates the magnitude of the magnetic field $B$.}
\end{figure*}

For 1D current sheets the problem of invariant conservation can be reduced to determining the rate of adiabatic invariant destruction (and thus $T_{AI}$) for the so-called transient particle trajectories \cite{Sonnerup71,Whipple86,BZ89}. Particles moving along transient trajectories carry the most significant current density in 1D current sheets \cite{Pritchett91, Burkhart92TCS,Sitnov00,Mingalev07}, and destruction of the corresponding adiabatic invariant may be interpreted as a destruction of the current sheet equilibrium \cite[see discussion in][]{Zelenyi02:grl,Zelenyi03}. The adiabatic invariant for transient particles experiences two random jumps per bounce period, when a particle approaches the current sheet and when a particle departs from the current sheet. Each individual jump is called {\it scattering}, with the same term also describing the cumulative effect of multiple jumps. The magnitude of these jumps is quite different for different current sheets. Therefore, the rate of the adiabatic invariant destruction (and correspondingly $T_{AI}$ and $T_{syst, model}$) are determined by a specific current sheet and the values of system parameters. Additionally, for some values of the parameteres, there are no separation of time-scales of motion, and there are no adiabatic invariants at all. In this study we aim to apply the machine learning approach to determine parametrical domains of existence of adiabatic invariant over observationally-relevant times $T_{Sim} \sim T_{syst}$ in three typical current sheets shown in Fig.~\ref{fig1}. 

Particle scattering and the corresponding adiabatic invariant destruction \citep[e.g.,][]{Chirikov79,Birmingham84,BZ89} are most effective when the minimum radius of curvature of magnetic field line, $R_c=B_nL/B_0$, is comparable to the maximum particle gyroradius, $\rho=v_0mc/eB_n$ ($v_0$, $m$, and $e$ are particle speed, mass, and charge, respectively). Thus, the ratio 
\[
\kappa = \sqrt{R_c/\rho}=(B_n/B_0)\sqrt{LeB_0/v_0mc}
\] 
is a key quantity defining the number of adiabatic invariants and the overall properties of particle dynamics for all three systems.

In the rest of this section we introduce three current sheets, and then look at them in some details. In Sect.~\ref{sec:discussion} we discuss obtained results of the number of invariants in the context of construction of plasma equilibria for realistic space plasma systems.

\subsection{Three current sheet models: overview}\label{subsec:00}

The first, and the simplest, configuration describes the compressional (with magnetic field magnitude variation) planar current sheet  (all field lines are within a plane) with $B_m=0$, see Fig.\ref{fig1}(a). Such current sheets are typical for planetary magnetotails \cite{Petrukovich15:ssr,Artemyev14:pss,Halekas06:currentsheets,DiBraccio15,Dubinin12,Poh17}, and can be found in the magnetopause around reconnection regions \cite{Gosling82,Dubinin02,Lavraud02}. The Hamiltonian for particles in such configurations is
\[
H = \frac{1}{{2m}}p_l^2  + \frac{1}{{2m}}p_n^2  + \frac{1}{{2m}}\left( {p_m  + \frac{{eB_n r_l }}{c} - \frac{{eB_0 }}{c}\frac{{r_n^2 }}{{2L}}} \right)^2 
\] 
We normalize coordinates and momentum as $z=r_n/\sqrt{Lv_0mc/eB_0}$, $x=r_l/\sqrt{Lv_0mc/eB_0}$, $p_x=p_l/mv_0$, $p_z=p_n/mv_0$, and energy and time as $\Ham = H/mv_0^2$, $t\to tv_0/\sqrt{Lv_0mc/eB_0}$. Constant $p_y = p_m/mv_0$ can be removed by shift the $x$-axis. A new Hamiltonian is:
\begin{equation}
    \Ham = \frac{1}{2}p_z^2  + \frac{1}{2}p_x^2  + \frac{1}{2}\left( {\kappa x - \frac{1}{2}z^2 } \right)^2 \label{eq:ham01}
\end{equation}
The value of Hamiltonian $\Ham$ can be set to $h=1/2$ by a proper choice of $v_0$. Particle scattering in such current sheets was extensively studied for $\kappa \ll 1$ regime \cite{Chen92,Burkhart91:Differentialmemory,Vainchtein05} and for $\kappa \geq 1$ regime \cite{Delcourt94:scattering, Delcourt96:choas,Shustov15,Lukin21:pop:df}. 

For strongly curved magnetic field lines, when $\kappa\ll 1$, particle scattering results in adiabatic invariant destruction over the time-scale $T_{AI} \sim \kappa^{-3}$ due to particle crossing the separatrix that is present on the phase plane of fast variables $(z,p_z)$ \cite[see details in][]{Neishtadt86,Cary86,BZ89}. In the opposite case of a large background $B_n$ field, when $\kappa \gg 1$, $T_{AI} \sim \kappa\exp\left(\kappa^2\right)$ \cite{Birmingham84}, that a typical accuracy of adiabatic invariant conservation for systems without a separatrix \cite{Cohen78,Chirikov79,Neishtadt00}. In Sect.~\ref{subsec:01} we investigate a full $\kappa$ range to determine parameters corresponding to one (total energy) and two (total energy and adiabatic invariant) conserved quantities in Hamiltonian (\ref{eq:ham01}).

The second configuration is encountered in current sheets in the solar wind and in the planetary magnetopause (see Fig.\ref{fig1}(b)). They are typically less compressional than the magnetotail current sheet, because they are embedded into strong shear $B_m$ fields \cite[e.g.,][]{Artemyev19:jgr:solarwind,Haaland20,Lukin20}. These fields are described by a vector potential $A_l(r_n)$, and for constant $B_m$ Hamiltonian is
\begin{eqnarray}
    \Ham &=& \frac{1}{2}p_z^2  + \frac{1}{2}\left(p_x-sz\right)^2  + \frac{1}{2}\left( {\kappa x - \frac{1}{2}z^2 } \right)^2 \nonumber \\ 
    \label{eq:ham02} \\
s &=& (B_m/B_0)\sqrt{LeB_0/v_0mc} \nonumber 
\end{eqnarray}
Particle scattering in current sheets with $B_m=const$ was studied in, e.g., \cite{Karimabadi90,Chapman98,Delcourt00,Artemyev13:npg}. 
For weak $B_m$ ($s<1$), there is a separatrix on the fast $(z,p_z)$ phase plane. Thus $T_{AI} \sim \kappa^{-3}$ for $\kappa\ll 1$ and $T_{AI}\sim \kappa\exp(\kappa^2)$ for $\kappa\gg 1$. For sufficiently strong fields, $s\geq 1$, there is no separatrix, and thus $T_{AI}$ is expected to be much longer \cite{Artemyev13:npg2}. In Sect.~\ref{subsec:02} we determine the domain in the parametric space $(\kappa, s)$ where system contained two invariants (total energy and adiabatic invariant).

The third configuration defines totally compressionless current sheets (so-called force-free current sheets, see Fig.\ref{fig1}(c) and \cite{Harrison09:prl,Allanson15,Neukirch18}), which are widely observed in the solar wind \cite{Artemyev19:jgr:solarwind,Neukirch20} and magnetopause \cite{Panov08,Haaland20,Lukin20}, and are characterized by $B_m=\sqrt{B_0^2-B_l^2}$. Such current sheets describe rotational discontinuities with strongly curved magnetic field lines. For the typical thin current sheets, $L\approx mv_0c/eB_0$, Hamiltonian is
 \begin{equation}
    \Ham = \frac{1}{2}p_z^2  + \frac{1}{2}\left(p_x-z+\frac{1}{3}z^3\right)^2  + \frac{1}{2}\left( {\kappa x - \frac{1}{2}z^2 } \right)^2 \label{eq:ham03}
\end{equation}
Dynamics of particle in such current sheets were studied in, e.g., \cite{Malara21,Artemyev20:pre}, where a very strong destruction of adiabatic invariants were found. This destruction is due to {\it geometrical jumps} of the adiabatic invariant \cite{Neishtadt&Treschev11,Artemyev14:pre:Ey&By}. In Sect.~\ref{subsec:03} we determine this effect as a function of $\kappa$. 

\subsection{Current sheet with $B_l(r_n)$, $B_n\ne 0$, and $B_m=0$}\label{subsec:01}

\begin{figure}
\centering
\includegraphics[width=0.45\textwidth]{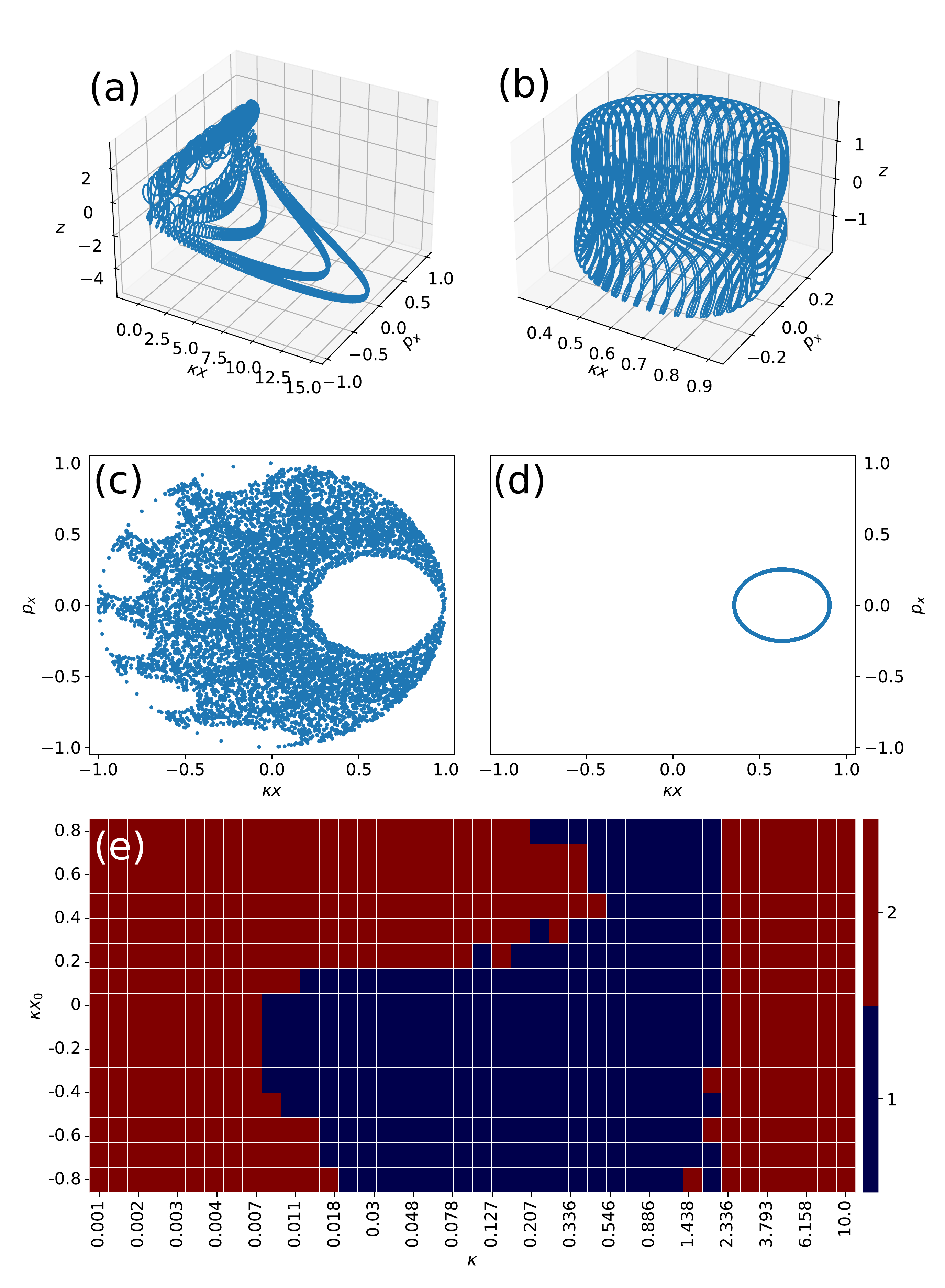}
\caption{
\label{fig2} (a)-(b) Typical particle trajectories in $(\kappa x, p_x, z)$ space for Hamiltonian system (\ref{eq:ham01}) for $\kappa=0.142$. Initial conditions were $\kappa x_0 = -0.35$ for (a) and $\kappa x_0 = 0.35$ for (b), (c)-(d) corresponding Poincar\'e maps for these two orbits, (e) number of invariants in the $(\kappa,\kappa x_0)$ plane.}
\end{figure}

We start the analysis from small $\kappa$, and then move to larger $\kappa$. Figures~\ref{fig2}(a, b) show two typical trajectories in the $(z,\kappa x, p_x)$ space for Hamiltonian (\ref{eq:ham01}) for $\kappa=0.142$ ($\kappa\ll 1$ regime). In the normalized variables, $p_x = \kappa y$, and thus Fig.~\ref{fig2}(a, b) effectively show orbits in 3D Cartesian coordinates. Particle dynamics defined by Hamiltonian (\ref{eq:ham01}) is a combination of fast oscillations in the $(z,p_z)$ plane and slow quasi-periodical motion in the $(\kappa x, p_x)$ plane. Panel (a) shows a so-called {\it transient} orbit that consists in two fragments: with and without $z=0$ crossing during one $(z,p_z)$ oscillation. Transient particles leave the vicinity of the plane $z=0$ and move along magnetic field lines to large $B_l$ values, where they are returned by the mirror force back. The right panel shows a so-called {\it trapped} orbit, that crosses $z=0$ twice each period of $(z,p_z)$ oscillations. Trapped particles spend all time around the $z=0$ plane. Averaging over fast $(z,p_z)$ oscillations yields an adiabatic invariant $I_z=(2\pi)^{-1}\oint{p_zdz}$ \cite{Sonnerup71,BZ86}. Transition between regimes with and without $z=0$ crossings results in the destruction of the $I_z$ conservation \cite{Neishtadt87,BZ89}. Figures \ref{fig2}(c, d) show the Poincar\'{e} maps confirming this difference between transient and trapped orbits. Conservation of $I_z$ reduces the number of degrees of freedom from $3$ to $2$, and thus the Poincar\'{e} section of the orbit by $z=0$ plane is a simple closed curve (Fig.~\ref{fig2}(d)). Destruction of $I_z$ makes particle orbits stochastic, and the Poincar\'{e} section of the orbit by the $z=0$ plane is a $2D$ set of points. The empty circle in the center of the Poincar\'{e} section for the transient orbit in Fig.~\ref{fig2}(c) is the part of the plane belonging to the trapped orbits. 

Transient and trapped orbits are separated in the parametrical $(\kappa x_0, \kappa)$ space, where $\kappa x_0$ is an initial particle position at $p_x=0$ ($z_0$ and $p_{z0}$ are defined in such a way that all particles have the same energy $h=1/2$). For $\kappa x_0 \in(-1, 0.1234)$, particles move along transient trajectories for which the destruction of $I_z$ is controlled by $\kappa$: $T_{AI} \sim \kappa^{-3}$ \cite{BZ89,Vainchtein05}. 

Up to $\kappa \sim 5\cdot10^{-3}$, $I_z$ is conserved on the time-scales with hundreds of $z=0$ crossings for each trajectory, and both trapped and transient orbits have two invariants (energy and $I_z$), see Fig.~\ref{fig2}(e). As $\kappa$ increases, the destruction rate of $I_z$ for transient particles with $\kappa x_0 \lesssim 0.1234$ also increases (and $T_{AI}$ decreases). At $\kappa \gtrsim 10^{-1}$, transient orbits have only one invariant (energy), whereas trapped orbits have two invariants. 

At $\kappa \gtrsim 0.3$ Hamiltonian system (\ref{eq:ham01}) loses the separation between fast $(z,p_z)$ and slow $(\kappa x, p_x)$ motions, and adiabatic invariant disappears for both trapped and transient orbits. Around $\kappa\sim 1$ there is a range of fully stochastic motion, where only energy is conserved. For $\kappa \gtrsim 2.5$ there is a new separation of time-scales: $(\kappa x, p_x)$ become fast variables and $(z,p_z)$ become slow variables. For fixed values of $(z,p_z)$, Hamiltonian (\ref{eq:ham01}) always has a single minimum as a function of $\kappa x$. Therefore, there is a single regime of the $(\kappa x, p_x)$ oscillations. For large $\kappa$, magnetic field lines near $z=0$ are along the $z$-axis (normal to the current sheet) and the adiabatic invariant $I_x=(2\pi)^{-1}\oint{p_x \; d\kappa x}$ is the magnetic moment, which is conserved with an exponential accuracy $\sim \exp(-\kappa^2)$ \cite{Birmingham84,Neishtadt00} for all $\kappa x_0$. Thus, for $\kappa>2.5$ all particle trajectories in the system have two invariants (energy and $I_x$) on all realistic time scales. 

Note that all threshold values of $\kappa$ depend on the time interval of the simulations, i.e. on the relation between the time-scale of $I_z$ or $I_x$ destruction ($T_{AI}$) and the interval of simulation ($T_{Sim}$).

\subsection{Current sheet with $B_l(r_n)$, $B_n\ne 0$ and constant $B_m\ne 0$}\label{subsec:02}
\begin{figure}
\centering
\includegraphics[width=0.45\textwidth]{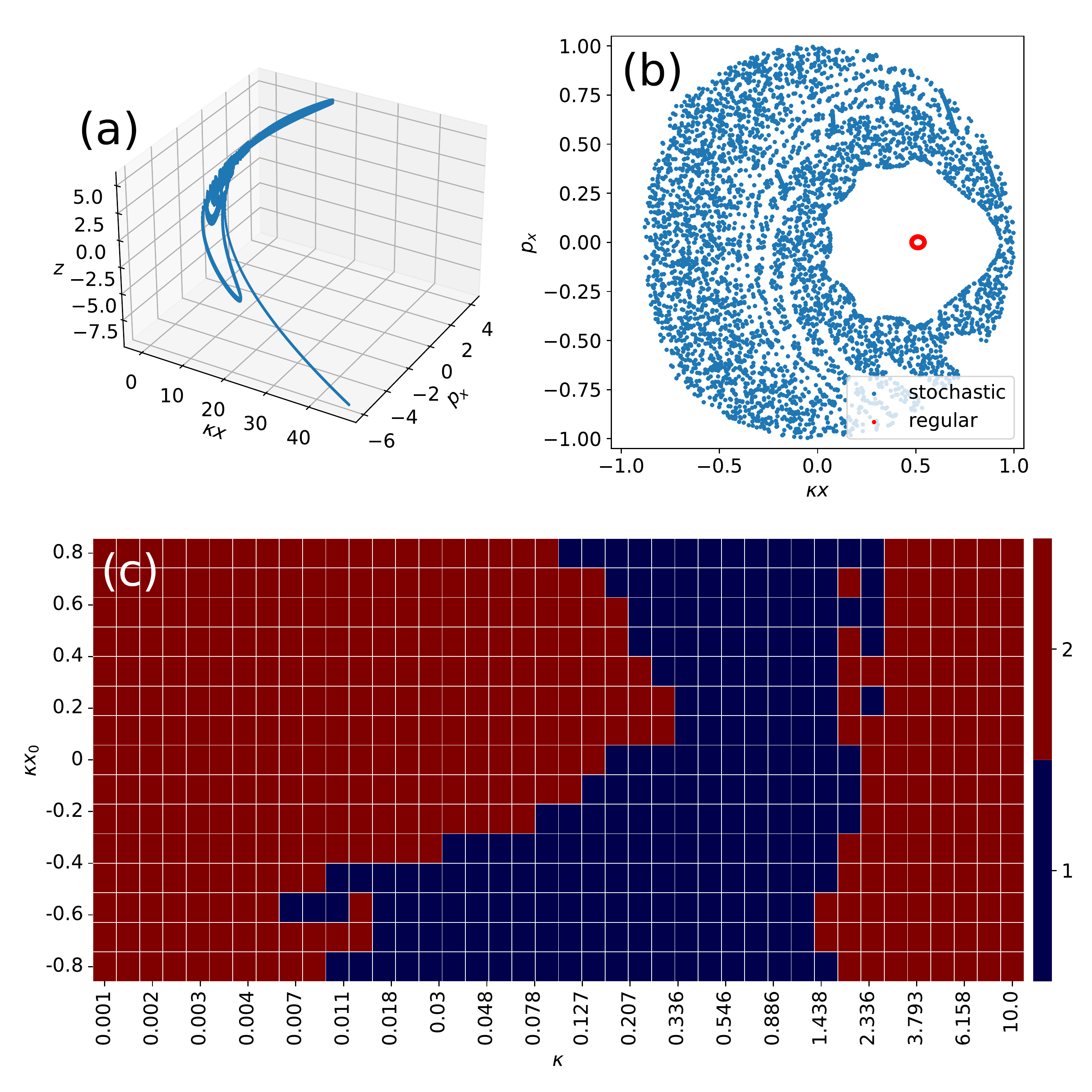}
\caption{
\label{fig3} (a) A typical transient trajectory in $(\kappa x, p_x, z)$ space for Hamiltonian (\ref{eq:ham02}) for $s=0.7, \kappa=0.142$ and $\kappa x_0 = -0.35$, (b) corresponding Poincar\'e maps for the orbit from Panel (a) (blue dots) and a trapped orbit (red circle), (c) number of invariants in the $(\kappa,\kappa x_0)$ plane.}
\end{figure}

Figure~\ref{fig3}(a) shows a typical transient orbit in the $(z,\kappa x, p_x)$ space for Hamiltonian (\ref{eq:ham02}) with $\kappa=0.142$. 
For sufficiently small $B_m$ magnitudes ($s<1$), there are well separated transient and trapped orbits. For transient orbits, in addition to the $I_z$ destruction due to changes of the type of motion (with and without $z=0$ crossings), there is a splitting of $I_z$ into several values: particles jump between different values of $I_z$ \cite{Artemyev13:npg}, called geometrical jumps of adiabatic invariant \cite{Neishtadt&Treschev11,Artemyev14:pre:Ey&By}. For symmetric magnetic fields, these jumps are reversible and by themselves do not contribute to the long-term $I_z$ destruction. However, in combination with the $I_z$ destruction with a rate of $\sim \kappa^3$ \cite{BZ89,Vainchtein05}, these jumps make transient orbits more stochastic \cite{Artemyev13:npg2,Zelenyi13:UFN}. Poincar\'e section in Fig.~\ref{fig3}(b) confirms that transient orbits (blue dots) are stochastic and possess only one invariant (energy), while trapped orbits (red circle) are regular and possess two invariants (energy and adiabatic invariant $I_z$). Separating such random and regular the Poincar\'e maps, we determine number invariants in the $(\kappa x_0, \kappa)$ plane. 

Figure~\ref{fig3}(c) shows that for small $\kappa<10^{-2}$ both transient and trapped orbits have two invariants for $s=0.7$. At $\kappa \sim 10^{-2}$ transient orbits (smaller $\kappa x_0$) loose one invariant. Comparison of Figs.~\ref{fig2}(e) and \ref{fig3}(c) shows that both for systems with $s=0$ and $s\ne 0$ there is a range near $\kappa \sim 1$ of fully stochastic motion with a single invariant (energy) for all orbits. The threshold $\kappa$ value, where $I_x$ appears, is slightly smaller for $s\ne 0$ than for $s=0$: two invariants for all $\kappa x_0$ exist for $\kappa >2$ at $s=0.7$ and for $\kappa >2.5$ at $s=0$. This is likely an effect of magnetic curvature decrease and corresponding weakening of the particle scattering and magnetic moment $I_x$ destruction \cite{Birmingham84,BZ89} with $s\ne 0$ (for $B_m \ne 0$).

\begin{figure}
\centering
\includegraphics[width=0.45\textwidth]{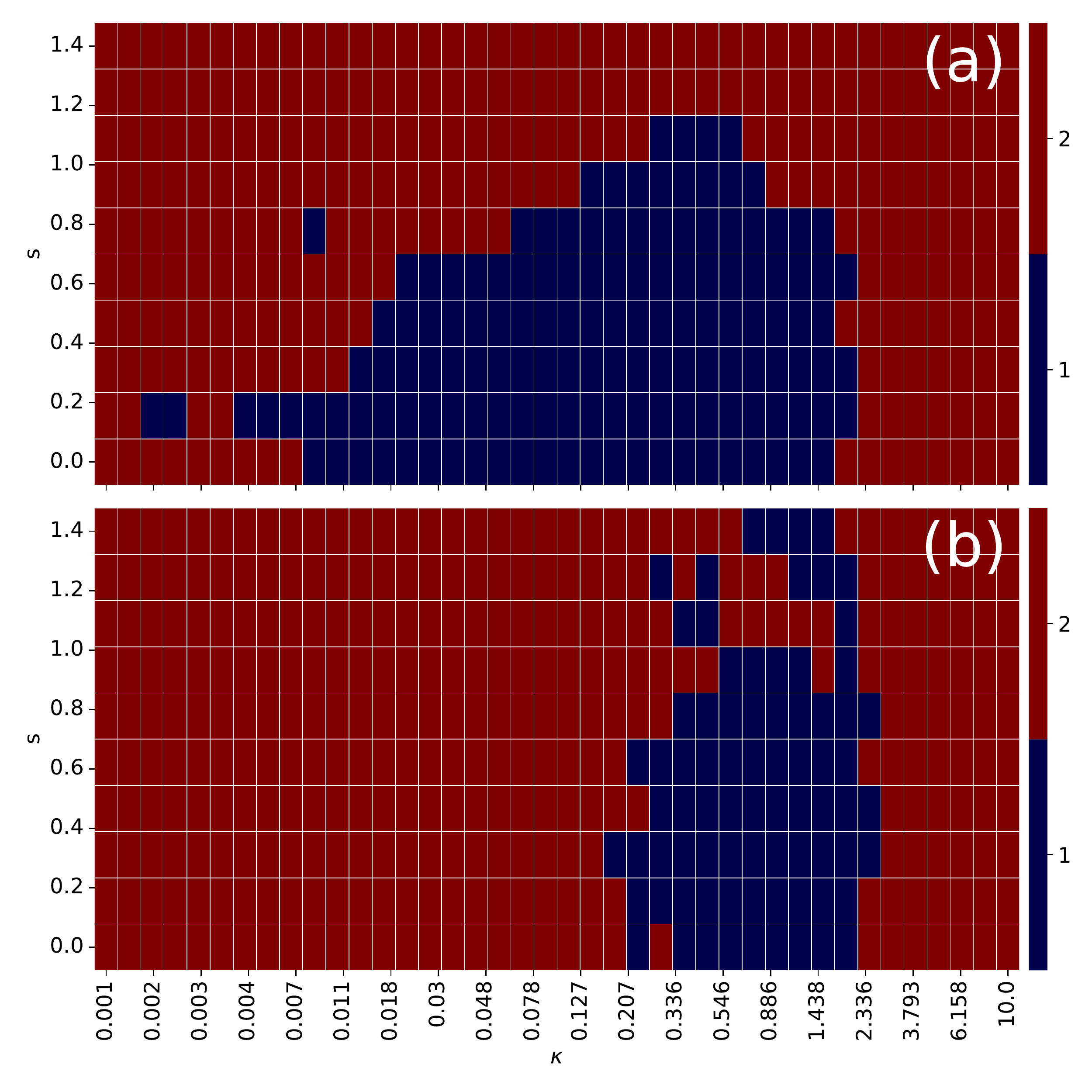}
\caption{
\label{fig4} Number of invariants in the $(\kappa,s)$ plane for two $\kappa x_0$. Top: $\kappa x=-0.35$ (transient orbit), bottom: $\kappa x=0.35$ (trapped orbit).}
\end{figure}

Figure \ref{fig4} shows the total number of invariants in the $(s,\kappa)$ parametric space. Two panels correspond to transient ($\kappa x_0 = -0.35$, top) and trapped ($\kappa x_0 = 0.35$, bottom) trajectories. The low-$s$ region ($s<1$) resembles Fig.~\ref{fig3}(c): when $\kappa<10^{-2}$ there are two invariants (energy and $I_z$) for trapped and transient orbits, when $\kappa \sim 10^{-1}$ the adiabatic invariant for transient orbits is destroyed, while there two invariants for trapped orbits, when $\kappa\sim 1$, there is a single invariant (energy) for trapped and transient orbits, when $\kappa>2.5$ there are two invariants (energy and $I_x$) for trapped and transient orbits (for $\kappa$ so large there is no real separation between trapped and transient orbits). The large-$s$ region ($s>1$) shows two invariants for all $\kappa$. This is the effect of magnetic radius of curvature increase (due to large $B_m$) and trajectories with and without $z=0$ crossing are similar. When $s>1$, the adiabatic invariant $I_z$ resembles the magnetic moment with a very weak destruction rate \cite{Artemyev13:npg2}.

\subsection{Current sheet with $B_l(r_n)$, $B_n\ne 0$ and $B_m(r_n)$}\label{subsec:03}
\begin{figure}
\centering
\includegraphics[width=0.45\textwidth]{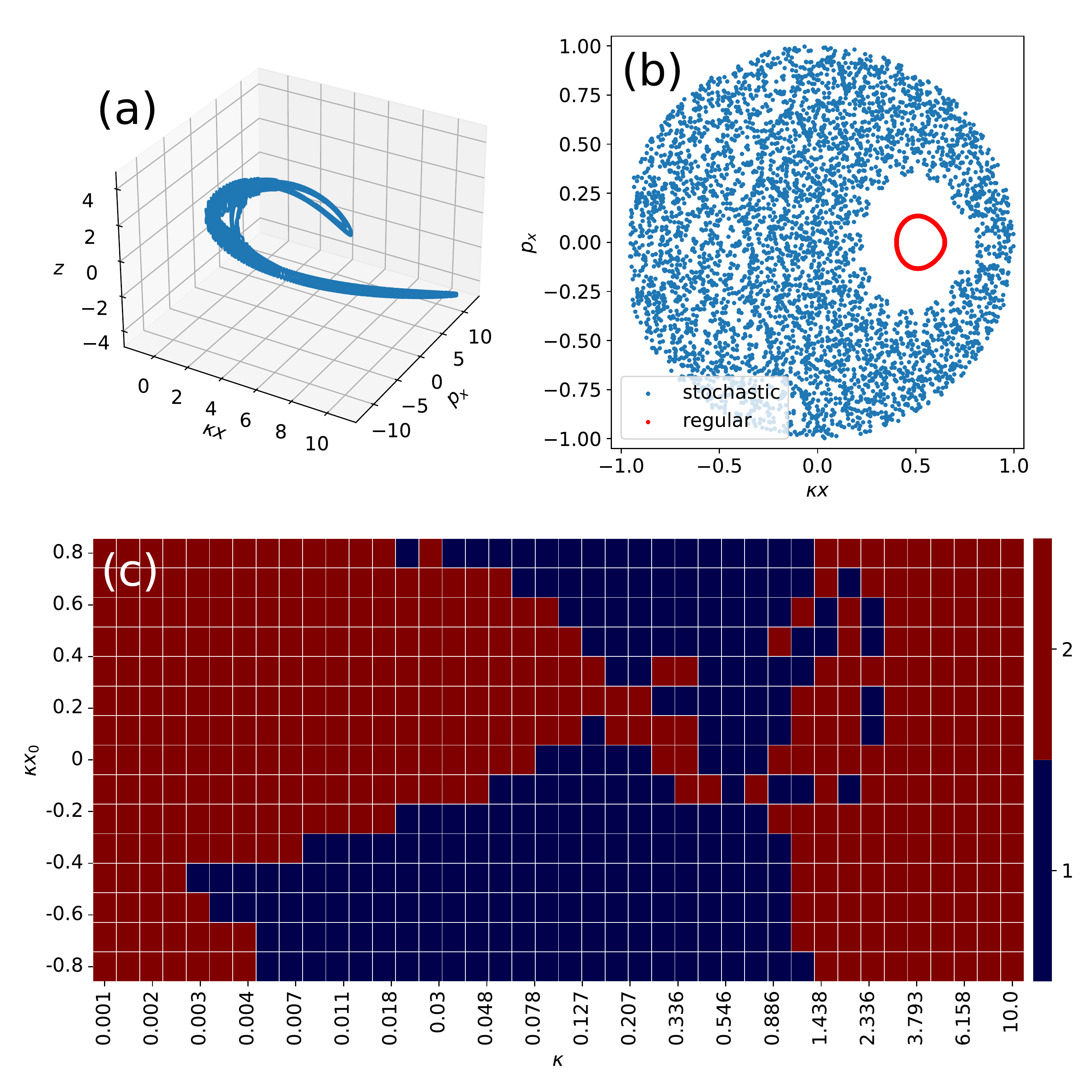}
\caption{
\label{fig5} (a) A typical trajectory in the $(\kappa x, p_x, z)$ space for Hamiltonian (\ref{eq:ham03}) for $\kappa=0.142$ and $\kappa x_0 = -0.35$; (b) corresponding Poincar\'e maps for the orbit from Panel (a) (blue dots) and a trapped orbit (red circle), (c) number of invariants in the $(\kappa,\kappa x_0)$ plane.}
\end{figure}

Figure~\ref{fig5}(a) shows a typical transient orbit in the $(z,\kappa x, p_x)$ space for Hamiltonian (\ref{eq:ham03}) that describes particle motion in curved magnetic field lines of force-free current sheets, typical for the solar wind and Earth's magnetopause (see Fig.~\ref{fig1}(c)). Poincar\'e section in Fig.~\ref{fig5}(b) shows an almost uniform distribution of transient trajectories. Transient particles experience strong scattering \cite{Artemyev20:pre,Malara21} due to geometrical jumps of adiabatic invariant. As magnetic field lines are not planar, this change occurs within a relatively small domain in the $(\kappa x, p_x)$ plane \cite{Artemyev20:pre}. Trapped particles do not experience the geometrical destruction of $I_z$, which explains the presence of regular curves in the Poincar\'e section (Fig.~\ref{fig5}(b)).

Comparison of Figs.~\ref{fig2}(e) and \ref{fig5}(c) shows several differences in particle dynamics between compressional and force-free current sheets. First, the $\kappa$ range corresponding to one invariant (energy) for transient orbits ($\kappa x_0<0.2$) is wider for force-free current sheet: $\kappa \in [3\cdot 10^{-4},2]$ for force-free versus $\kappa \in [5\cdot10^{-3},2]$ for compressional current sheets. Second, in force-free current sheets there is a range of $\kappa x_0$ between transient and trapped regions where particles have two invariants for all $\kappa$ values. There is no such population in compressional current sheets, where for $\kappa \sim 1$ the entire $\kappa x_0$ range is filled by stochastic orbits having only one invariant of motion (energy). The large-$\kappa$ boundary where the second invariant (likely $I_x$) is at $\kappa\sim 2$ for both current sheets. 

%

\section{Discussion and Conclusions} \label{sec:discussion}
In this study we applied a machine learning approach, AI Poincar\'e \cite{Liu&Tegmark21:ML}, to determine the number of invariants of charged particle motion in several current sheets. We considered three configurations most typical for planetary magnetopauses and magnetotails and for the solar wind. For all three configurations, the AI Poincar\'e determines parametrical domains with two invariants (energy and an adiabatic invariant of fast periodical oscillations) and those with a single invariant (energy). In the latter domains either the adiabatic invariant is destructed by particle scattering, or there are no separation of time-scales of motion, and thus there are no adiabatic invariants at all. We investigated the number of invariants for transient orbits, whose contribution dominates the current density \cite{Eastwood72,Burkhart92TCS,Pritchett92,Sitnov00,Mingalev07}), and for trapped orbits, that do not contribute to the current density in $B_n=const$ current sheets \cite{Pellat&Schmidt79}. In parametrical domains with a single invariant (energy), particle distribution functions do not support any  current density. Therefore, the existence of stationary current sheet models with $B_n=const$ (or, more general, the existence of stationary current sheets with $B_n=const$) is determined by the presence a population of transient particles with two invariants. Taking this into account, the most important conclusions of our study are:
\begin{enumerate}
    \item For all three considered current sheets, there are no additional (exact or approximate) invariants of charged particle motion besides the energy and one adiabatic invariant. Therefore plasma equilibria describing such current sheets should be constructed with the particle velocity distributions written as one or two-invariant distributions. There is no hidden symmetry allowing generalization of existing classes of equilibria \cite{Sitnov06,Zelenyi11PPR}.
    \item For all three considered current sheets, for strongly curved magnetic field lines ($\kappa \ll 1$) the parametrical domain with two invariants is limited to very small $\kappa\leq 5\cdot10^{-3}$ values. The shear $B_m$ magnetic field (both constant and with a peak at $z=0$ plane) increases the $\kappa$ threshold for two invariant domain near the boundary between transient and trapped particles (around $\kappa x_0\sim 0$), i.e. there is a current carrying particle population supporting the current sheets even with $\kappa\sim 10^{-1}$ (see Figs. \ref{fig3} and \ref{fig5}).
    \item For compressional current sheets (see Fig.~\ref{fig1}(a)), there is a large parametrical domain with a single invariant for transient particles: $\kappa \in[5\cdot10^{-3}, 2]$, that quite typical values of $\kappa$ for observed current sheets. Thus, there is almost no transient particles carrying current in stationary 1D current sheets for such $\kappa$. For observed current sheets, this range of $\kappa$ includes all reasonable energies ($m v_0^2/2 \propto \kappa^{-4}$) of ion populations. Current sheets observed within this range should be nonstationary \cite{Zelenyi03, Zelenyi02:grl} and are quickly destroyed or evolve into essentially 2D structures (where the current is carried by trapped ions drifting in $\partial B_n/\partial r_l$, see \cite{SB02,Birn04}). 
\end{enumerate}

Obtained results show that 1D stationary current sheets with $B_n\ne0$ have at best two invariants and do not have a sufficiently wide parametrical range filled by transient particles with two invariants, although some small domains can be found for $B_m \ne 0$ configurations. As in realistic space plasma systems fluctuations of the background system parameters ($B_n$, $B_m$) alter boundaries of such domains (if they do exist), the very existence of fully stationary 1D current sheets is questionable. Most likely, observed 1D current sheets are dynamical equilibria, where the inflow of transient particles compensates scattering of resident particles (scattering leads to adiabatic invariant destruction and current density reduction). Such inflow can be supported by external sources of transient particles (see, e.g., discussion in \cite{Maha93,Maha05,Zelenyi06:beamlet}) or scattering of trapped particles into the parametrical domain corresponding to transient particles (e.g., due to magnetic field line curvature \cite{Zelenyi90, Burkhart95} or small-scale transient fluctuations of magnetic field \cite{Veltri98,Greco00,Greco02,Artemyev16:pre}). Existing stationary 1D current sheet models are essentially based on the assumption of eternal adiabaticity of transient particle dynamics \cite{Sitnov03,Sitnov06,Sitnov&Merkin16,Zelenyi06, Zelenyi11PPR}. Further modification of such models with inclusion of dynamical effects (scattering, inflows and trapped/transient population exchange) should describe life-times of observed current sheets.

\begin{acknowledgments}
L.A.S., A.A.V., P.A.A. acknowledge support from Russian Science Foundation through grant No. 19-12-00313.
\end{acknowledgments}

%

\end{document}